\documentclass{aa}
\usepackage{latexsym,graphics,exscale,graphicx}
\usepackage{pictexwd,rotating,amssymb,nicefrac}
\usepackage{booktabs}

\begin{document}
%
%
   \title{The merger stage of the ultra-luminous infrared galaxy  
IRAS\,03158+4227}

   \author{H. Meusinger\,$^{1,\star}$,
         \ B. Stecklum\,$^{1,\star}$,
         \ C. Theis\,$^{2}$,
         \ J. Brunzendorf\,$^{1,}$
\thanks{
Visiting Astronomer, German-Spanish Astronomical Centre, Calar Alto,
operated by the Max-Planck-Institute for Astronomy, Heidelberg, jointly 
with the Spanish National Comission for Astronomy}
}


   \institute{
$^{1}$ Th\"uringer Landessternwarte Tautenburg, 07778 Tautenburg,
Germany\\
$^{2}$ Institut f\"ur Theoretische Physik und Astrophysik der
       Universit\"at Kiel, Olshausenstr. 40, 24098 Kiel, Germany}

   \date{Received ; accepted }

\abstract{
We examine the merger stage of IRAS\,03158+4227, one of the most
luminous ULIRGs from the IRAS 2\,Jy sample. Deep optical images 
are presented along with high-resolution NIR images and
optical low-resolution spectra. IRAS\,03158+4227 is confirmed as a 
component of an equal-luminosity binary galaxy with a nuclei 
separation of $47\,h_{75}^{-1}$\,kpc.  
A long lopsided tail  emanating from the companion, which 
harbours an active nucleus, is the most prominent peculiar feature 
of the binary.  The results of numerical 
simulations admit the interpretation of this structure as a product
of the tidal interaction between the two components. 
If the infrared-activity of IRAS\,03158+4227 is also dynamically triggered 
by this process, this would implicate that the ULIRG phenomenon is not 
restricted to the final stage of a binary merger. Alternatively, the 
system may be a multiple merger where the partner(s) has/have already
coalesced.
}
   
\authorrunning{Meusinger, Stecklum, Theis, Brunzendorf}
\titlerunning{IRAS\,03158+4227}
\maketitle

%
%
%
\section{Introduction}
%
%

The characteristics of galaxies that emit a substantial amount of their
bolometric luminosity in the far infared (FIR) has been a matter
of debate since their discovery (e.g., Rieke \& Low \cite{Rie72};
Joseph \& Wright \cite{Jos85}; Soifer et al. \cite{Soi87}).
Particular interest is focussed on the class of
ultra-luminous infrared galaxies (ULIRGs),  i.e. galaxies
with  quasar-like infrared luminosities of $L_{\rm IR} \ge 10^{12}L_{\sun}$
(e.g., Sanders \& Mirabel \cite{San96}; Kennicutt \cite{Ken96};
Genzel et al. \cite{Gen98}; Rigopoulou et al. \cite{Rig99};
Scoville et al. \cite{Sco00}).
Nuclear starbursts and/or AGN activity,
dynamically triggered by gravitational interactions, are thought to be
the energy sources in the cores of ULIRGs. Absorption and re-emission
in the FIR is expected to be a consequence of thermalization of radiation 
by large masses of dust grains.

It has been demonstrated by a large number of  studies
that a high fraction of luminous and ultraluminous
infrared galaxies show morphological peculiarities,
such as tidal debris and double nuclei, indicative of
gravitaional perturbations.
For instance, all ULIRGs of the original IRAS bright galaxy survey
(BGS; Sanders et al. \cite{San88}) show indications of strong
gravitational interaction or merging.
Published values for the interacting/merging fraction
among the ULIRGs from various studies cover the range
from 50\% to about 100\% (e.g.,
Auri\`ere et al. \cite{Aur96};
Clements et al. \cite{Cle96};
Murphy et al. \cite{Mur96};
Duc et al. \cite{Duc97};
Borne et al. \cite{Bor00}).
Murphy et al. have analysed near-infrared (NIR) and visual
images for 46 luminous infrared galaxies with
$L_{\rm IR} > 8.5\,10^{11}\,L_{\sun}$ from the IRAS 2\,Jy sample.
After combining their sample with the BGS, they find that 95\%
of the galaxies in the combined sample show evidence for
current or past interactions. 
There are only three ULIRGs in the Murphy et al. sample that do not,
to the limits of the images given there, show indications for
interactions. Among them is IRAS\,03158+4227, one of the 
most luminous ULIRGs 
($\log\,L_{\rm IR}/L_{\sun} = 12.55,\ f_{60}/f_{100} = 0.95$).
This seems astonishing since the fraction of perturbed systems among the
FIR-bright galaxies is known to increase with $L_{\rm IR}$ (e.g.,
Sanders \& Mirabel \cite{San96}, and references therein).

On the numerical side, simulations clearly show that the major
morphological features
observed in many peculiar galaxies are explained as due to tidal forces
during galaxy encounters  (e.g.,
Toomre \& Toomre \cite{Too72};
Barnes \& Hernquist \cite{Bar92};
Bekki \& Noguchi \cite{Bek94};
Spoke \cite{Spo97};
Mihos et al. \cite{Mih98}).
Self-consistent models of tidally disturbed galaxies indicate
large gas concentrations in the centres due to strong and sudden
gaseous inflow (Negroponte \& White \cite{Neg83};
Noguchi \cite{Nog91};
Barnes \& Hernquist \cite{Bar91}, \cite{Bar96};
Mihos \& Hernquist \cite{Mih96}).
The models predict that merger-driven
gas-dynamics and associated star formation may result in spectacular
starbursts (e.g.  Noguchi \& Ishibashi \cite{Nog86};
Mihos \& Hernquist \cite{Mih96}),
although there is considerable uncertainty about the treatment of star
formation and of the feedback from young stars in the simulations.
Galaxy merger and the ULIRG phenomenon were tied together
in the models by Mihos \& Hernquist (\cite{Mih96}).

ULIRGs are widely claimed to represent the final stages of
merging galaxies. Murphy et al.(\cite{Mur96}) give projected 
linear separations of less than a few kpc for a
large fraction of their sample. 
In the case of  IRAS\,03158+4227, Murphy et al. 
noticed a nearby, resolved component which they described
as ``not apparently interacting with the primary galaxy''.
IRAS\,03158+4227 was therefore not included by these authors 
in their sample of double-nuclei systems and was subsequently
considered a single system where the nuclei separation must
be smaller than the available resolution of $0\farcs8$.
The projected linear distance to the nucleus of the galaxy mentioned
by Murphy et al. amounts to $18''$ corresponding to a projected
linear separation of 47\,kpc (throughout this paper we adopt an 
Einstein-de Sitter comology with  $H_0 = 75$\,km\,s$^{-1}$\,Mpc$^{-1}$). 
There are only a few 2\,Jy ULIRGs with nuclear distances larger 
than 10\,kpc and only  one other system (IRAS\,14394+5332) 
with a separation of about 50\,kpc.

In this paper, we present deep optical imaging,
high-resolution (adaptive optics) NIR imaging, and optical spectroscopy
of the remarkable ULIRG IRAS03158+4227.
The observational data (imaging and spectroscopy) are described in
Section\,2. The results are presented in Sect.\,3 and are discussed
in the context of simulations in Sect.\,4.
Finally, conclusions are given in Sect.\,5.

%
%
%
\section{Observations and data reduction}
%

Most of the observations presented here were taken at the
German-Spanish Astronomical Centre on Calar Alto, Spain.
A summary of relevant data for all observations is given in Table\,\ref{obs}.

\begin{table}[th]   
\caption{Log of observations.
}
\begin{tabular}{rrrr}
\toprule
filter/    & $t_{\rm exp}$   &  instrument &  epoch\\
grism      & [s]         &             &        \\
\hline
J          &    360      & ALFA/Omega-Cass       & Sep\,98\\
H          &    180      & ALFA/Omega-Cass       & Sep\,98\\
K$'$       &     90      & ALFA/Omega-Cass       & Sep\,98\\
-          &    700      & CAFOS       & Jul\,98\\
B          & 4\,560      & CAFOS       & Jul\,98,00\\
R          & 3\,700      & CAFOS       & Jul\,99,00\\
I          &    820      & CAFOS       & Jul\,00\\
B\,400     & 1\,200 (G1) & CAFOS       & Jul\,99\\
B\,400     & 3\,000 (G2) & CAFOS       & Jul\,99\\
R          & 1\,200      & TLS Schmidt & Jan\,99\\
I          &    960      & TLS Schmidt & Dec\,99\\
\bottomrule
\label{obs}
\end{tabular}
\end{table}

Deep direct optical imaging was performed with the focal reducer
camera CAFOS at the 2.2\,m telescope in the B, R, and I band and without
filter. CAFOS was equipped with a SITe CCD with a scale of $0\farcs5$/pixel.
The conditions were always photometric with a seeing of
typically about $1''$. Unfortunately, IRAS\,03158+4227 is located only
about $2'$ away from the bright foreground star HD\,20489 (V=8.6)
and 7\farcs4 from the USNO-2 star No.275-02191303 (B=16.2, R=14.9). 
Therefore, we took several sets of relatively  short exposures
(typically between 100 and 300\,s) to avoid saturation effects of 
the CCD. The total  integration time amounts to 2.8\,hours.
MIDAS standard algorithms were applied for the data reduction. 
The combined image was PSF-deconvolved using the 
Lucy-Richardson method (MIDAS procedure deconvolve/flucy).

High-resolution imaging in the J, H, and K$'$ bands was performed using
the adaptive optics system ALFA in combination with the Omega-Cass camera
(Hippler et al. \cite{Hip98}) at the 3.5\,m Calar Alto telescope.
Omega-Cass utilizes a 1024$\times$1024 HAWAII detector and was operated
at the pixel scale of 0\farcs08. Since there is no sufficiently bright
star close to IRAS~03158+4227 which could be used for wavefront sensing
and with the laser guide star being not operational at the time of the
observations, we corrected the static aberrations of the telescope on a
nearby star and imaged the target with the deformable mirror ``frozen''.
Although this procedure does not yield diffraction-limited resolution,
it improved the image quality considerably, leading to sub-arcsecond 
resolution. Two adjacent fields were observed for deriving the sky frames.
During the data processing, the images were rebinned in order to enhance
the signal-to-noise ratio which led to a final pixel scale (from the 
astrometric solution) of 0\farcs155. After correction for flat field and 
bad pixels, the images were filtered using the wavelet algorithm of Pantin \&
Starck (\cite{Pan96}) to
minimize noise amplification in the subsequent Richardson-Lucy deconvolution.
The USNO-2 star No.275-02191303 served as PSF reference for the deconvolution.
JHK$'$ photometry was derived from the non-deconvolved images and tied to the
JHK$'$ magnitudes of the PSF star according to its entry in the 2MASS Second
Incremental Release Point Source Catalog (Cutrie et al. \cite{Cut00}).
The derived fluxes (Fig.\,\ref{SED}) refer to a synthetic aperture
of 7\farcs75 diameter.  The photometric error amounts to 0.06 mag.
In order to assess the separation of a possible double nucleus, the FWHM of
the images of the ULIRG in the three filters were compared to those of stars in
the field. The average stellar FWHMs derived from Gaussian fits amount
to 0\farcs60, 0\farcs57, and 0\farcs58 for J, H, and K$'$, respectively.
The FWHMs of the ULIRG are 0\farcs74, 0\farcs70, and 0\farcs71.
This leads to beam-deconvolved sizes of 0\farcs44, 0\farcs40, and
0\farcs40 for
the angular extent of the emitting core region of the ULIRG.

\begin{figure}[bhtp]   
\vspace{9cm}
\includegraphics{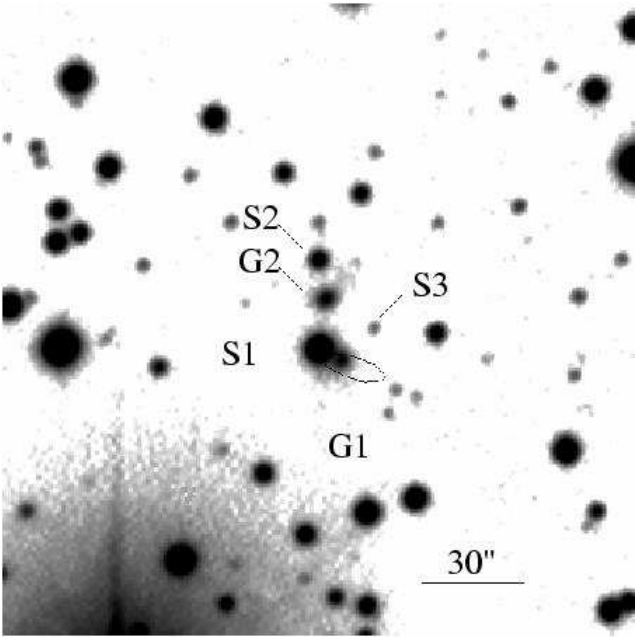}
 \caption{IRAS error ellipse of IRAS\,03158+4227. Coaddition of
R- and I-band images taken with the Tautenburg Schmidt camera.
The size of the image is
$200''\times 200''$; N is up, E is left. S1 to S3
designate stars, G1 and G2 indicate galaxies.}
  \label{ellipse}
\end{figure}

Low-resolution spectra of both IRAS\,03158+4227
and its nearest neighbour galaxy
were taken with CAFOS equipped with the grism B\,400
which is suitable for the wavelength range $\lambda = 3\,200...8\,000$\,{\AA}.
With a slit width of $1\farcs2$ the spectral resolution is about
20\,{\AA}.

Finally, we observed the field of IRAS\,03158+4227 with the
Tautenburg Schmidt telescope at moderate seeing of about $2''$.
The Schmidt camera was equipped with a 2\,k$\times$2\,k SITe CCD
with pixel size of 24\,$\mu$m$\times$24\,$\mu$m which yields
a field size of 40'$\times$40'. These images were used only to
evaluate the large-scale environment of IRAS\,03158+4227.

%
%
%
\section{Results}
%
%

Figure\,\ref{ellipse} shows a moderately deep optical image of the  
field around IRAS\,03158+4227.
The data for the IRAS error ellipse were taken from the IRAS
Point Source Catalogue, the coordinates for the objects on the
optical image
are from the astrometric solution for the field of the Perseus
cluster of galaxies (Brunzendorf \& Meusinger \cite{Bru99}). The
IRAS error ellipse clearly overlaps with the optical image of a faint
galaxy (G1) with $z = 0.13$
(NASA Extragalactic Database, NED). At the distance of the galaxy,
the size of the field shown in Fig.\,\ref{ellipse}
corresponds to about 0.5\,Mpc$\times$0.5\,Mpc.
There is no sign for a dense galaxy cluster or a rich group
of galaxies around IRAS\,0315+4227. Several small and faint galaxy
images are seen in the field, but a substantial fraction may
be dwarf galaxies of the Perseus cluster in the foreground.

\begin{figure}[bhtp]   
\vspace{6cm}
\includegraphics{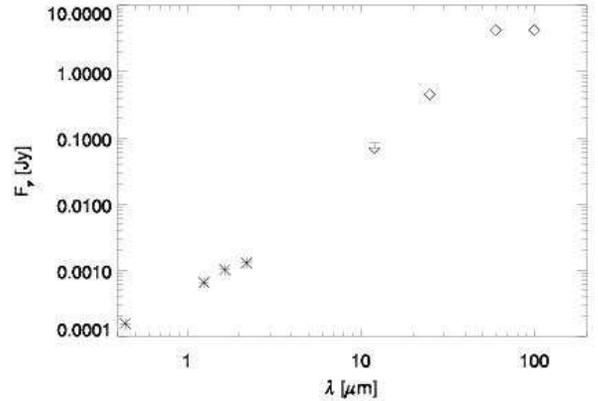}
 \caption{Spectral energy distribution (SED) of IRAS~03158+4227. Asterisks
denote our measurements while diamonds mark the IRAS fluxes. (The data point
at 12\,$\mu$m represents an upper limit.)
}
  \label{SED}
\end{figure}

Murphy et al. (\cite{Mur96}) presented a K-band image taken
with the Palomar 200\,inch telescope at a seeing of $0\farcs8$
that does neither show evidence for a neighbour galaxy outshined
by the brighter foreground star S1 on a less resolved image, nor
for a double nucleus
of G1. As mentioned in  Sect.\,1, IRAS\,03158+4227 was
assumed to be a late merger with nuclei separation of less than
$0\farcs8$, even though it was noted by Murphy et al.
(\cite{Mur96}) that there is a galaxy (G2) at a distance of
$18''$. The two galaxies G1 and G2 have approximately
the same apparent magnitudes; after calibrating our B-band image
using stars from the USNO-A2.0 Catalogue (Monet et al. \cite{Mon98})
we find $B =18.6$ and 18.4 for G1 and G2, respectively.
This corresponds to $M_{\rm B} \approx -21.4$ and $-21.6$
(for $A_{\rm B} = 0.8$\,mag from the NED\footnote{
The NASA/IPAC Extragalactic Database (NED) is operated by the Jet
Propulsion Laboratory, California Institute of Technology, under contract
with the National Aeronautics and Space Administration.}, 
and a $k$-correction
$k_{\rm B} = 0.5$\,mag from Coleman et al. \cite{Col80})
and to a very strong FIR excess of
$\log L_{\rm FIR}/L_{\rm B} = 2.3$ for G1
(see e.g., Soifer et al. \cite{Soi87}
for the definitions of $L_{\rm FIR}$ and $L_{\rm B}$).

The spectral energy distribution (SED) of IRAS~03158+4227
is shown in Fig.\,\ref{SED}. The SED is typical for ULIRGs, with
the bulk of the luminosity radiated at far-infrared wavelengths.
In addition to the data points shown in Fig.\,\ref{SED}, we note
that IRAS~03158+4227 is identified with a radio continuum
source in the NVSS (Condon et al. \cite{Con98}). The flux density
of 12.4\,mJy at 1.4\,GHz corresponds to a radio-to-infrared flux
ratio (for definition see Helou et al. \cite{Hel85}) $q=2.64$.
This is in agreement with the well-known correlation between
the flux densities in
the infrared and the radio continuum (see Sanders \& Mirabel \cite{San96},
and references therein) where $\langle q \rangle \approx 2.35$ for
most of the luminous infrared galaxies in the BGS, but is larger
for the galaxies with highest infrared luminosities.
(For instance, Helou et al. found $q=2.60$ for Arp\,220.)

\begin{figure*}[htbp]
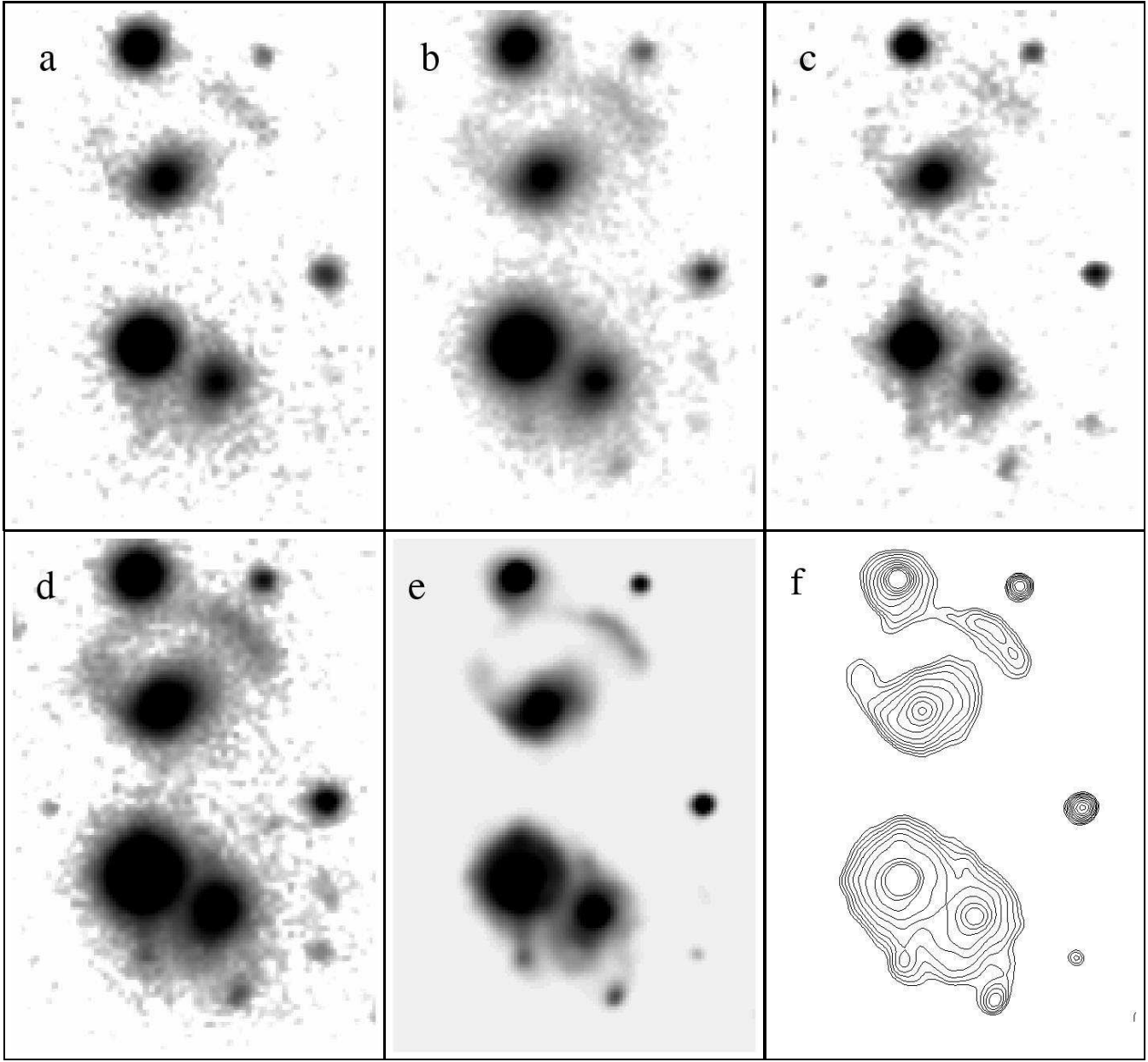
  
 \begin{tabbing}
 \fbox{\resizebox{5.7cm}{8.01cm}{\includegraphics{meusinger1671.f3a}}} \hfill \=
 \hspace{-0.22cm}
 \fbox{\resizebox{5.7cm}{8.01cm}{\includegraphics{meusinger1671.f3b}}} \hfill \=
 \hspace{-0.25cm}
 \fbox{\resizebox{5.7cm}{8.01cm}{\includegraphics{meusinger1671.f3c}}} \hfill\\
 \hspace{-0.1cm}
 \fbox{\resizebox{5.7cm}{8.01cm}{\includegraphics{meusinger1671.f3d}}}\hfill\=
 \hspace{-0.12cm}
 \fbox{\resizebox{5.7cm}{8.01cm}{\includegraphics{meusinger1671.f3e}}} \hfill \=
 \hspace{-0.25cm}
 \fbox{\resizebox{5.7cm}{8.01cm}{\includegraphics{meusinger1671.f3f}}} \hfill\=
\end{tabbing}
\caption{Optical images of IRAS\,03158+4227: {\bf a} B-band; {\bf b} R-band; 
{\bf c} I-band; {\bf d} co-addition of all optical images; 
{\bf e} the same image as {\bf d}, 
but  after Lucy-Richardson deconvolution with 10 iteration steps; 
{\bf f} the contour plot of image {\bf e} where successive contour lines
correspond to a factor 3 in intensity. (The unfiltered image shows the same
structures as the images {\bf a} to {\bf c} and is not displayed here).
The size of each image is $30''\times 42''$, N is up, E is left.
}
\label{optical}
\end{figure*}

The optical images (Fig.\,\ref{optical}) clearly reveal
that G2 has a faint, but very extended material arm on the side
opposite to G1. This feature is seen in all optical bands, and
the Lucy-Richardson-deconvolved image shows that it
is quite narrow as expected for tidal tails. In the optical bands,
the arm is about 3\,mag fainter than the main body of G2.
The surface brightness of the brightest part of the tail
is estimated to $\mu_{\rm B} \approx 25$\,mag\,arcsec$^{-2}$,
the projected linear extent is about 70\,kpc. Illustrative examples
for similar tidal structures with one dominating arm can be
found among the  Arp (\cite{Arp66}) and Arp \& Madore (\cite{Arp87})
galaxies (e.g., Arp\,107, 110, 129, 173, 252, 255, AM\,2350-302,
AM\,0552-324, AM\,0427-475).
For the host galaxy of the ULIRG itself, tidal signatures are
much weaker, although there appears to be some
fuzz around G1. Several brighter spots may represent tidal
debris, perhaps including huge star forming regions; but
clearly, none of these spots is bright enough to be considered
a second nucleus of G1. Unfortunately, it can not be excluded that
faint morphological features of G1 are hidden behind 
the bright star S1.

\begin{figure}[htbp]
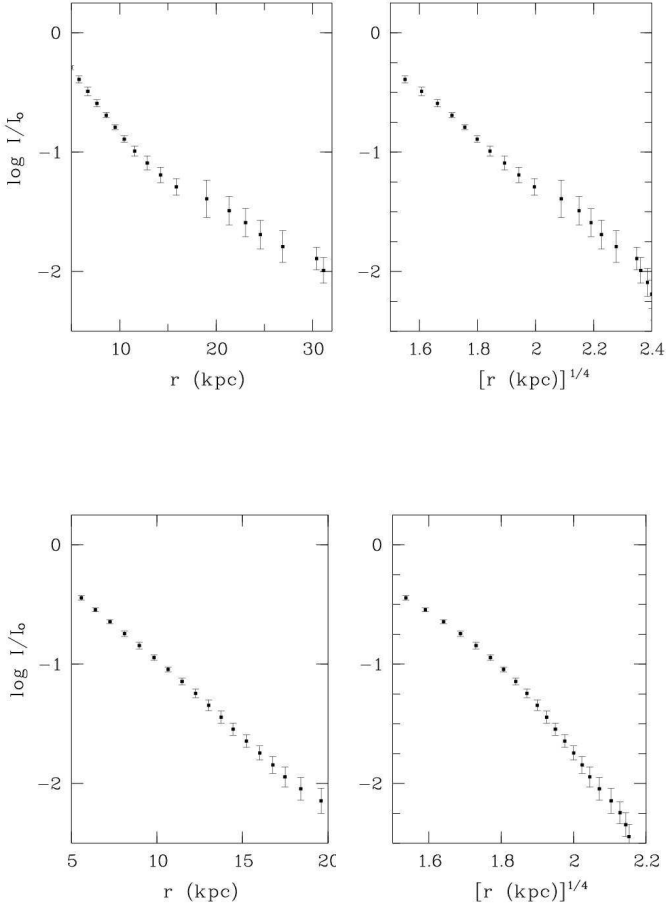
  
 \begin{tabbing}
\resizebox{4.7cm}{6.7cm}{\includegraphics{meusinger1671.f4a}} \hfill\=
\hspace{-0.4cm}
\resizebox{4.5cm}{6.7cm}{\includegraphics{meusinger1671.f4b}} \hfill \\
\resizebox{4.7cm}{6.7cm}{\includegraphics{meusinger1671.f4c}} \hfill\=
\hspace{-0.4cm}
\resizebox{4.7cm}{6.7cm}{\includegraphics{meusinger1671.f4d}} \hfill \\
\end{tabbing}
\vspace{-0.6cm}
\caption{Intensity profiles of G1 (top) and G2 (bottom) derived from
the R-band image by means of the MIDAS procedure fit/ell3. 
$I_0$ is the central intensity. The angular interval where the image 
of G1 is affected by the star S1 was excluded from the profile analysis.
}
\label{profile}
\end{figure}

The radial intensity profiles of G1 and G2 in the R-band are
shown in Fig.\,\ref{profile}. G1 may be approximated by two components 
with different scalelengths. Alternatively, it can be 
classified as an ``E-like'' ULIRG: the radial surface density profile 
is reasonably fit by a deVaucouleurs $r^{1/4}$-law over the range 
$R \approx$\,4 \ldots 30\,kpc. (The innermost 1\farcs5 were excluded 
from the analysis.) According to Sanders et al. (\cite{San00}), 
about one third of the ULIRGs from the 1\,Jy sample
are classified as ``E-like''. Contrary to G1, the profile of G2 is
better approximated by an exponential law which suggests
a ``disk-like'' structure.

\begin{figure*}[bhtp]   
\vspace{8.2cm}
\includegraphics{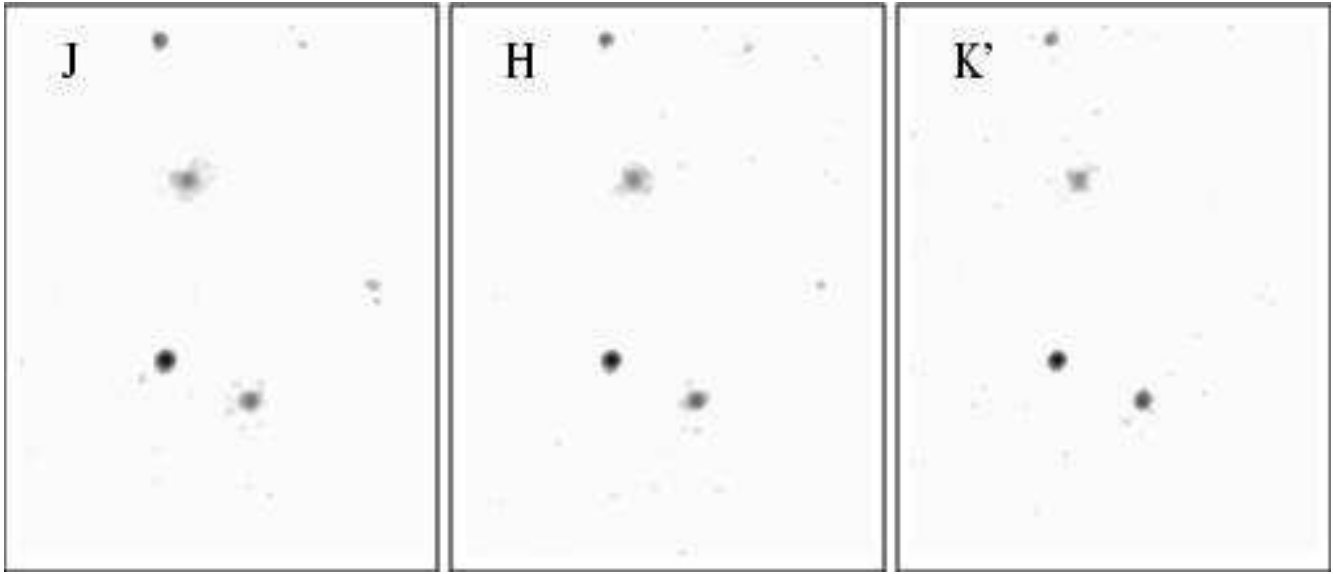}
 \caption{High-resolution images of IRAS\,03158+4228 in the J, H, and K$'$
bands. The scale and the size are the same as in Fig.\,\ref{optical}.
}
  \label{K-band}
\end{figure*}

The images in the NIR-bands are shown in Fig.\,\ref{K-band}.
These high-resolution images were taken to search for point sources
indicating either a close double (or multiple) nucleus with small
separations or a nucleus outshined by the bright star S1 in the optical
image, rather than to evaluate extended structures of low-surface
brightness. For such an aim, observations at longer wavelengths are needed,
since the morphology of the central parts of ULIRGs is strongly affected
by dust obscuration. As expected (cf. Sanders et al. \cite{San00}),
the light distribution in K$'$ is very compact for the galaxy G1
which appears as a point source. Figure\,\ref{K-band} does not reveal 
any sign for a
double nucleus in G1 or for a nucleus close to S1 down to the resolution
(beam-deconvolved size) of 0\farcs40. Of course, there remains the
possibility of an incidental superposition of the image of S1 and
a further nucleus, but the probability for such a configuration is very
low.

\begin{figure}[bhtp]   
\vspace{11.8cm}
\includegraphics{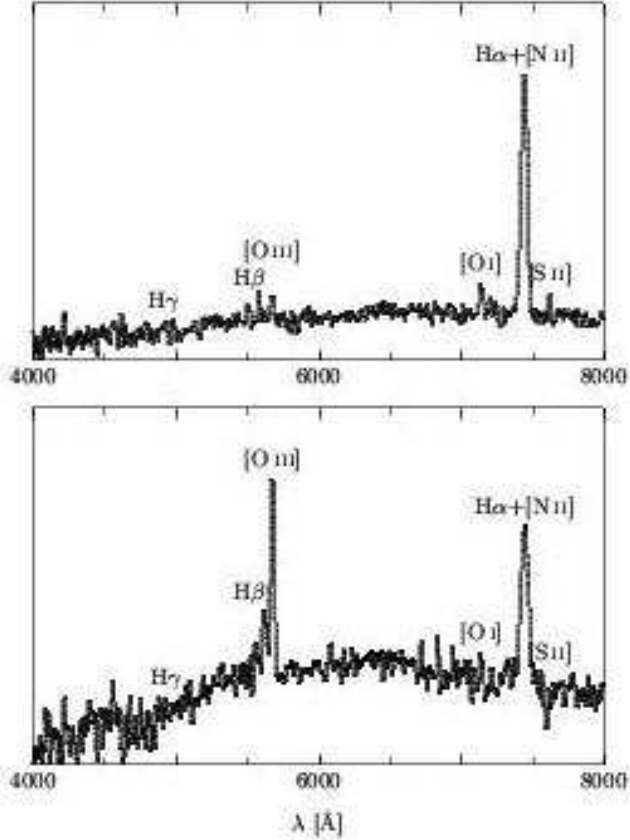}
\caption{Optical low-dispersion spectrum (not flux-calibrated, observer frame) 
of the ULIRG IRAS03158+4227 (G1, top) and of the galaxy G2 (bottom).}
\label{spec}
\end{figure}

The optical spectra of G1 and G2 (Fig.\,\ref{spec}) show strong emission
lines; the equivalent widths (EWs) are listed in Table\,\ref{lines}. 
A radial velocity difference between G1 and G2 of $240$\,km\,s$^{-1}$ 
is estimated, but the uncertainty due to the low spectral resolution
is quite large (about $\pm200$\,km\,s$^{-1}$).

IRAS\,03158+4227 has sometimes been classified as a Seyfert\,2,
whereby strong absorption was invoked as the reason for
the absence of indications of nuclear activity in
the hard X rays (Risaliti et al. \cite{Ris00}).
In the spectrum from Fig.\,\ref{spec}, we do not see clear-cut
evidence for an AGN in IRAS\,03158+4227; in particular the
conventional diagnostic EW ratio [\ion{O}{iii}]5007/H$\beta$
is too low for a Seyfert nucleus. However, the signal-to-noise
ratio is low for these
lines, and, owing to the low dispersion, we
are not able to resolve H$\alpha+$[\ion{N}{II}]6584.
The low intensity of the emission lines in the blue
part of the spectrum is probably best explained as due to
strong internal dust absorption and resultant
reddening, though the effect of the underlying stellar
absorption is difficult to estimate. If H$\beta$ is
significantly affected by stellar absorption of an
older, A-type burst population, the intrinsic [\ion{O}{iii}]5007/H$\beta$
ratio would be even less compatible with a Seyfert spectrum.
On the other hand, the diagnostic line ratios of G2
are more consistent with an AGN.
It should be noted that the K$'$-band image of G1 is stronger 
concetrated than that of G2 (Fig.\,\ref{K-band}) as seems 
typical for ULIRGs as compared with AGN hosts
(Sanders et al. \cite{San00}).

\begin{table}[th]   
\caption{Equivalent widths EW (in units of {\AA}) of the emission
lines measured for the galaxies G1 and G2.
}
\begin{tabular}{lrr}
\toprule
                                    & G1             &  G2     \\
\hline
EW(H$\beta$)                        & $3.5\pm0.4$    & $2\pm2$ \\
EW([\ion{O}{iii}]5007)              & $8.0\pm0.4$    & $59\pm1$ \\
EW([\ion{O}{i}]6300)                & $8.0\pm0.4$    & $0.1\pm0.2$ \\
EW(H$\alpha$+[\ion{N}{ii}]6584)     & $118.7\pm0.2$  & $124.0\pm1.0$ \\
EW([\ion{S}{ii}]6717,6731)          & $3.7\pm0.3$    & $1\pm1$\\
\bottomrule
\label{lines}
\end{tabular}
\end{table}

%
%
%
\section{Discussion}
%
%

The long, curved tail of G2 seen in Fig.\,\ref{optical}, along with
the fact that G1 and G2 have nearly the same redshift, may be
taken as an indication for the gravitational interaction of G1 and G2.
Such tails of escaping debris from the far side of a victim disk are
well known indicators of the encounter of nearly equal-mass spiral galaxies
(e.g. Toomre \& Toomre \cite{Too72}; Schombert et al. \cite{Sch90}).
On the other hand, if the ULIRG activity was triggered by this
interaction, the large projected distance between these two
galaxies is surprising.
Murphy et al. (\cite{Mur96}) suggested that ULIRGs with large ($>10$\,kpc)
nuclear separation may represent triple merger with a third, undetected
nucleus from a previous encounter or, alternatively, that the ULIRG
phenomenon can occur in an early phase of interaction. Below, we briefly
discuss IRAS\,03158+5228 in the light of these two scenarios.

Arguments in favour of the multiple merger scenario were derived
from the properties of elliptical galaxies (e.g.,
Barnes \cite{Bar84}; Mamon \cite{Mam87}; Schweizer \cite{Sch89};
Weil \& Hernquist \cite{Wei96}), from the dynamical diversity of
ULIRGs (Borne et al. \cite{Bor00}; Cui et al. \cite{Cui01}), and
from detailed studies of  individual galaxies (Taniguchi \& Shioya
\cite{Tan98}; Lipari et al. \cite{Lip00}). Multiple encounter and
merger are suggested to occur naturally in compact groups
of galaxies (Barnes \cite{Bar89}; Hickson \cite{Hic97}; Borne et al.
\cite{Bor00}; Bekki \cite{Bek01}).
It seems likely that a fraction of ULIRGs is triggered
by such a process. Borne et al. (\cite{Bor00}) and Cui et al.
(\cite{Cui01}) considered the appearance of double or multiple nuclei
as a key-test for the multiple merger origin and derived
percentages of 20\% and 17\%, respectively, of multi-nuclei ULIRGs.
The fraction of ULIRGs
triggered by multiple merger is certainly  larger than the fraction
of multi-nuclei systems since a multiple nucleus is expected
to evolve on a short timescale to a double nucleus and finally
to a single nucleus. Unfortunately, this method is faced with serious
difficulties which can lead to an overestimation of multi-nuclei systems:
the morphology of
the central regions of ULIRGs has the tendency to be strongly
affected by dust obscuration effects and by the appearance of regions
of intense star formation on a scale of kpc or sub-kpc. Further,
the studies mentioned above did not identify real interacting
members with spectroscopic observations. Following Bekki (\cite{Bek01}),
it seems fair to say that  the fraction of ULIRGs formed
by multiple merging is still highly uncertain.

Dinh-V-Trung et al. (\cite{Din01}) have studied the six systems with
nuclear separations larger than 20\,kpc among the ULIRGs from the
complete 1\,Jy sample (Kim \& Sanders \cite{Kim98}).
The optical and K$'$-band imaging observations and optical spectra
suggest the multiple merger scenario only for one of those ULIRGs, 
IRAS\,14394+5332.
It can not be excluded that IRAS\,03158+4227 is a multiple merger like
IRAS\,14394+5332. Indeed, the morphologies of these two systems show some
similarities. Though we do not find evidence for a close double nucleus,
IRAS\,03158+4227 might be in a more advanced stage, where
two nuclei of G1 have coalesced and the inner region
is already well relaxed as seems indicated by the radial luminosity
profile of (Fig.\,\ref{profile}).

How can the nuclear activity of G2 be matched by such a scenario?
Nuclear activity is not unusual in ULIRG-systems, though the active 
nuclei are mostly located in the hosts of the ULIRG itself. However, 
there is one system (IRAS\,17028+5817) among the widely separated pairs 
studied by Dinh-V-Trung et al. where the spectrum of the ULIRG's host is
of \ion{H}{II}-type whereas the companion has a LINER-type spectrum.
We can not exclude that G2 is also a late merger. 
However, its disk-like structure admits a variant
of a multiple merger where the ULIRG activity was triggered by a past 
merger and the AGN in G2 by  the present interaction between G1 and G2.

The simulations by Bekki (\cite{Bek01})
have demonstrated that multiple merger can trigger repetitive
starbursts with a star formation rate comparable to ULIRGs. However, the
discussion by Bekki suggests that very intense starbursts with an
amplitude of $\ga 10^2$\,M$_{\odot}$/yr are not likely in such an
environment. As was stressed already in the
Introduction, IRAS\,03158+4227 is one of the most luminous ULIRGs
from the 2\,Jy sample. According to the relation derived by 
Clements et al. (\cite{Cle96}), the 60\,$\mu$m flux transforms into a
huge star formation rate of about 2\,10$^3$\,M$_{\odot}$/yr,
i.e. much higher than what seems possible in compact groups.

As an alternative to the multiple merger scenario, it seems tempting to 
speculate that the activities in the centres of G1 and G2 were triggered 
by the same process, namely an interaction of G1 and G2. 
Liu \& Kennicutt (\cite{Liu95}, their Fig.\,4) discussed the empirical
distribution of the equivalent widths of the H$\alpha$+[\ion{N}{ii}]
line for different merger morphological types. The EWs measured for
G1 and G2 (Table\,\ref{lines}) are in better agreement with
Liu \& Kennicutt's morphology type 3 (= systems of two disk galaxies)
than with type (2 = advanced merger) which appear to be single.
Moreover, according to its infrared colour index $f_{25}/f_{60}<0.2$,
IRAS\,03158+4228 belongs to the group of ``cool'' ULIRGs
which are characterized as major merger with prominent extended tidal
structures and resolved double nuclei rather than by small ($<2.5$\,kpc)
nuclei separation systems (Surace et al. \cite{Sur00}).

The simulations by Mihos \& Hernquist (\cite{Mih96})
have demonstrated that disk/bulge/halo systems  with dense central
bulges experience strongest gaseous infall and star formation activity
in the final stages of coalescence when they are within a few kpc
of one another. Their disk/halo models without dense bulges, on the
other hand, are most active in earlier phases of merging
when the galaxies are separated by tens of kpc. At the beginning of
the first starburst phase, the snapshots of the disk/halo merger
models by Mihos \& Hernquist (their Figs.\,11 and 12) show a
remarkable similarity with the few morphological details seen in
IRAS\,03158+4227: one galaxy (hereafter: g1) is more concentrated,
especially the gas and the young stars, with knots and short arms
whereas the most prominent feature of its interaction partner
(hereafter: g2) is an extended curved tail at the opposite side.
During the next time steps, when the SFR reaches its maximum,
the bridge between g1 and g2 becomes weaker and
g1 becomes more concentrated.

The long lopsided tail of G2 is the only visible morphologically peculiar 
feature of the system. It is therefore important for the understanding
of the merger stage of IRAS\,03158+4227 to know whether this structure
can be due to the tidal interaction with G1. Since it is not possible 
to follow the evolution of the extended tidal structures in the snapshots 
shown by Mihos \& Hernquist we performed a small series of 
{\it restricted N-body} simulations like those in Toomre \& Toomre (1972).
The main idea of this method is to derive the orbits of both galaxies
from the corresponding two-body problem, e.g.\, by solving a Kepler problem,
if the galaxies are treated as point masses. Using these orbits the
time-dependent potential at each point is given by a superposition of the
two galactic potentials. Stars are treated as test particles, by this
reducing the classical $N$-body problem to $N$ single-body problems.
(Details of the applied code are described in Theis \& Kohle (\cite{The01})).
The main advantages of this method are a fast computation and a high spatial
resolution. However, the method is not self-consistent, because
effects of self-gravity (like fragmentation or dynamical friction) are
neglected. Anyway, comparisons between self-consistent and restricted
N-body calculations demonstrated in several cases a good agreement, provided
the encounters are not too strong and/or the duration of the simulated stage
is not very long. Therefore, and because there are not many constraints 
from observations, the restricted $N$-body simulations should be a
good starting point. The results from the present simulations  
are considered however indicative rather than conclusive.

\begin{figure*}[htbp]
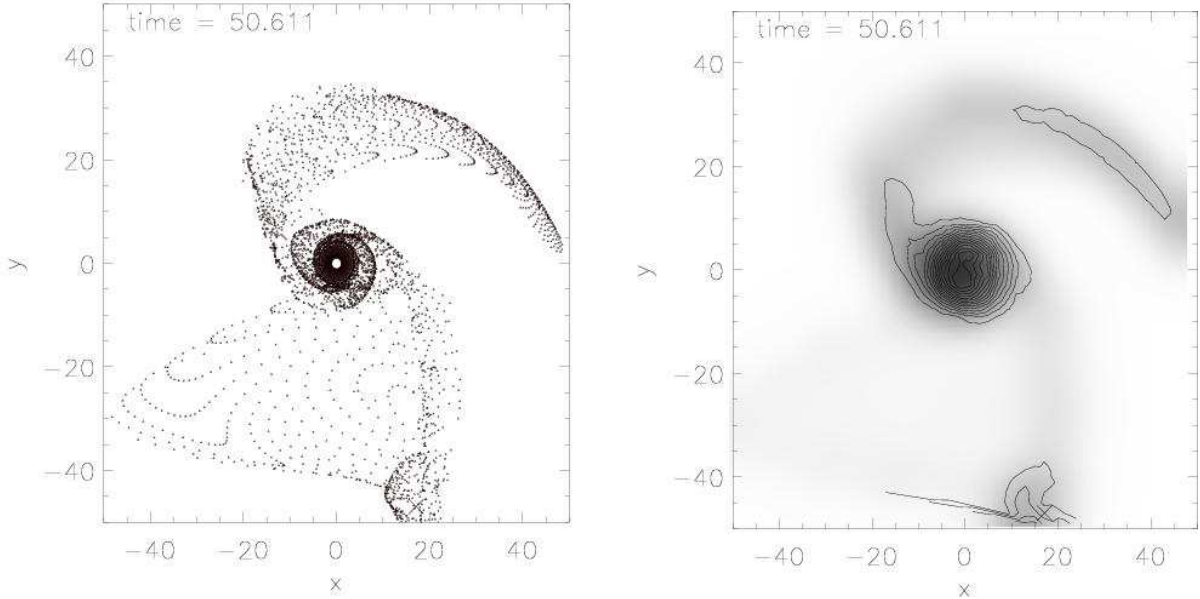
 
\vspace{0.5cm}
\vspace{8cm}
\includegraphics{meusinger1671.f7a}
\includegraphics{meusinger1671.f7b}
\caption{Simulation of the gravitational perturbation of the galaxy 
G2 by the galaxy G1. Both galaxies have the same dynamical mass of
10$^{12}$\,M$_{\odot}$.
Left: projected distribution of the mass particles from G2. 
G1 is marked by the cross at the bottom. 
The length is measured in kpc and the time unit is 1.49 Myr. No test
particles have been used inside the central kpc.
Right: blurred intensity plot of the image from the left hand side,
overlaid by a contour plot; the lowest 
intensity contour approximately corresponds to the detection 
threshold in the optical images of IRAS\,03158+4227.  
}
\label{simulations}
\end{figure*}

Figure\,\ref{simulations} shows the result of a parabolic encounter with an
orbit inclination of 60$^\circ$, a minimum distance of 15 kpc (reached at
$t=0$), a line-of-sight-velocity of 200 km\,s$^{-1}$, and a final projected
distance of 50 kpc. The total (dynamical) mass of each galaxy is
$10^{12}$M$_{\odot}$.  Since we are mainly interested in the tidal shape
of G2, we resolved only G2 in test particles. Motivated by the observed
radial intensity profile, we distributed the 
test particles in an exponential disk with a
scale length of 5 kpc and a cut-off radius at 15 kpc. For simplicity we
assumed the disk to be coplanar with the plane of sky.
The qualitative agreement between the optical image and the simulations
is very good:
after $7.5 \cdot 10^7 \, \mbox{yr}$ a lopsided
structure has been formed. This gives rise to an arc-like structure north-west
of G2 and a small spur emanating north-east  from the main body
of G2. In the simulations, these features are the dense parts of a tidal tail
which seems also to exist in the observations (e.g.\ lower-right panel of
Fig.\,\ref{optical}). The bridge
connecting G1 and G2 may be too weak to be clearly detected.
Increasing the line-of-sight velocity to 300 km\,s$^{-1}$ yields very
similar results.

The assumption that IRAS\,03158+4227 is triggered by the interaction 
between G1 and G2 implicates that the ULIRG phenomenon is 
not restricted to late binary merger stages. Such an 
interpretation is supported by further indications.
Rigopoulou et al.
(\cite{Rig99}) reported a lack of any correlation between
the stage of merger, measured by the separation of nuclei, and the
infrared luminosity in an unbiased sample of 62 ULIRGs. Further,
there is no trend for increased ULIRG activity in systems with
more centrally concentrated
H$\alpha$ emission (Mihos \& Bothun \cite{MiBo98}), and
also the total mass of molecular gas in ULIRGs is not
related to the linear separation (Gao \& Solomon \cite{Gao99};
Rigopoulou et al. \cite{Rig99}).
Finally, Dinh-V-Trung et al. (\cite{Din01}) present evidence for
IRAS\,23327+2913 to be hosted by a non-disturbed spiral-like galaxy
which may be interpreted as an early stage of merging.

%
%
%
\section{Conclusions}
%
%

Deep optical images, spectra and high-resolution NIR images 
are presented for IRAS\,03158+4227, one of the most luminous ULIRGs
of the 2\,Jy sample. The host galaxy, G1, is identified with a component 
of a binary of nearly equal-luminosity giant galaxies separated 
by $47\,h_{75}^{-1}$\,kpc. The companion galaxy, G2, harbours an 
active nucleus and has an extended, curved tail. This tail may  
be interpreted as a tidal structure induced by the gravitational 
interaction with the host galaxy of the ULIRG. The high-resolution
NIR images do not reveal any sign for a close double nucleus down
to a (resolution-limited) nuclei separation of 1\,kpc.

ULIRGs in such widely separated systems may be explained by a
multiple merger, like IRAS\,14394+532 (Dinh-V-Trung et al.
\cite{Din01}), where the interaction of
two components has already reached an advanced stage with a very
small separation of nuclei. Indeed, the optical luminosity profile of 
G1 is reasonably fit by a de\,Vaucouleurs law. 
The profile of G2 is more compatible with a disk-like structure. 
Hence it seems likely, in the multi-merger scenario, that the ULIRG in G1
was triggered by a past merger and the AGN in G2 by the present
interaction between G1 and G2. If the nuclear activity of G2 is 
also due to a (late) merger, the system IRAS\,03158+4227 would 
represent the case of the gravitational interaction of two advanced 
merger. Such scenarios would implicate that multiple merger in compact
groups can trigger very intense starbursts. 

As an alternative explanation, both the ULIRG IRAS\,03158+4227
and the nuclear activity in its neighbour galaxy might be dynamically
triggered by the same process, namely the
gravitational interaction of these two galaxies.
In this case, IRAS\,03158+4227 would represent an early stage of
binary tidal interaction, and the interpretation of ULIRGs as final
merger stages may need to be reexamined.

\begin{acknowledgements}

This research is based on observations made with the
3.5m telescope and the 2.2m telescope
of the German-Spanish Astronomical Centre, Calar Alto, Spain.
We acknowledge financial support from the Deutsche
Forschungsgemeinschaft under grants Me\,1350/3,5,8,14 and
Ste\,605/15. This research has made use of the NASA/IPAC Extragalactic
Database (NED) which is operated by the Jet
Propulsion Laboratory, California Institute of Technology, under
contract with the National Aeronautics and Space Administration.

\end{acknowledgements}


{}

\end{document}